\documentclass[11pt, twoside, eqno]{article}
\usepackage{amssymb}
\usepackage{mathrsfs}
\usepackage{amsmath}
\usepackage{amsfonts}
\usepackage{amsthm}
\usepackage{subfigure}
\usepackage{graphicx}
\usepackage{booktabs}
\usepackage{setspace}
\usepackage{float}

\allowdisplaybreaks \numberwithin{equation}{section}

\def\hs{\hspace{0.5cm}}

\newtheorem{thm}{\noindent\rm\bf Theorem}[section]
\newtheorem{obs}{\noindent\rm\bf Observation}

\newtheorem{lem}{\noindent\rm\bf Lemma}
\begin{document}
%\doublespacing
\baselineskip=15pt
\renewcommand{\arraystretch}{2}
\arraycolsep=1pt
\title{\bf \Large Zero-Correlation Linear Cryptanalysis of Reduced-round MISTY1
\footnotetext{\hspace{-0.6cm} ${ }^{*}$ Corresponding authors. \\
   E-mail addresses: nlwt8988@gmail.com. }
\author{\vspace{-0.1cm}\bf Wentan Yi$^{*}$ and Shaozhen Chen  \\
\vspace{-0.5cm}\small\it State Key Laboratory of Mathematical Engineering and Advanced Computing,\\
\small\it Zhengzhou 450001, China }}
\date{}

\maketitle

\begin{center}
\begin{minipage}{15.2cm}
\small{\bf Abstract.}  The MISTY1 algorithm, proposed by Matsui in FSE 1997, is a block cipher
with a 64-bit block size and a 128-bit key size.  It was recommended by the European NESSIE project and the CRYPTREC project, and became one RFC in 2002 and an ISO standard in 2005, respectively.

% Up to now, MISTY1 has attracted extensive attention and interests, and its security has
%been analysed against a wide range of cryptanalytic techniques. However, its security evaluation
%against the recent zero-correlation linear attacks is still lacking.
In this paper, we first investigate the properties of the FL linear function and identify $2^{32}$ subkey-dependent zero-correlation linear approximations over 5-round MISTY1 with 3 FL layers. Furthermore, some observations on the FL, FO and FI functions are founded and based upon those observations, we select $2^7$ subkey-dependent zero-correlation linear approximations and then, propose the zero-correlation linear attacks on 7-round MISTY1 with 4 FL layers. Besides, for the case without FL layers, $2^7$ zero-correlation linear approximations over 5-round MISTY1 are employed to the analysis of 7-round MISTY1.

\quad\, The zero-correlation linear attack on the 7-round with 4 FL layers needs about $2^{119.5}$ encryptions with $2^{62.9}$ known plaintexts and $2^{61}$ memory bytes. For the attack on 7-round without FL layers, the data complexity is about $2^{63.9}$ known plaintexts, the time complexity is about $2^{81}$  encryptions and the memory requirements are about $2^{93}$ bytes. Both have lower time complexity than previous attacks.

\medskip

\end{minipage}
\end{center}

\section{\large\bf Introduction}

MISTY1 is a block cipher designed by Matsui\cite{M1} in FSE 1997.
Since selected as one of Japanese e-Government standard ciphers by the CRYPTREC project in 2002,  MISTY1 became
widely deployed in Japan.
Outside of Japan,  MISTY1 was one of the final portfolio of the NESSIE-recommended ciphers\cite{N}, and approved
as one RFC \cite{M2} in 2000  and as an ISO standard\cite{I} in 2005, respectively. Besides, MISTY1 was selected as
the blueprint of the  GSM/3G block cipher KASUMI\cite{T}, which is one of the most widely used block ciphers in the world.
For those reasons, it is very important to understand the security offered by MISTY1.
\begin{table}
\caption{Summary of the attacks on MISTY1}
\centering
\scriptsize
\begin{tabular}{ccccccc}
\hline\noalign{\smallskip}
Attack Type & Rounds & FL Layer& Date Complexity & Time  Complexity & Source\\
\noalign{\smallskip}\hline\noalign{\smallskip}
HOD&7 &$3$&$2^{54.1 }$CP & $2^{120.7}$Enc & \cite{TSSK} \\
\noalign{\smallskip}\hline\noalign{\smallskip}
IA &6 &$4$&$2^{32}$CP & $2^{126.09}$Enc& \cite{SL}  \\
\noalign{\smallskip}\hline\noalign{\smallskip}
ID &6&None& $2^{54}$CP & $2^{61}$Enc & \cite{K2}  \\
ID & 7 &None & $2^{50.2}$CP & $2^{114.1}$Enc & \cite{LKKD}  \\
ID & 7 &None & $2^{55}$CP & $2^{92.2}$Enc & \cite{JL}  \\
ID & 6 &$4$ & $2^{52.5}$CP & $2^{112.4}$Enc & \cite{JL}  \\
ID &7&3& $2^{58}$KP & $2^{124.4}$Enc & \cite{JL}  \\
\noalign{\smallskip}\hline\noalign{\smallskip}
MZCL&7& None&$2^{63.9}$KP&$2^{81}$Enc & Sect.[5]\\
MZCL&7&4&$2^{62.9}$KP&$2^{118}$Enc & Sect.[4]\\
\noalign{\smallskip}\hline
\end{tabular}

HOD: Higher-Order Differential;
IA: Integral Attack;
ID: Impossible Differential;
MZCL: Multidimensional Zero-Correlation Linear; CP: Chosen Plaintext; KP: Known
Plaintext;  Enc: Encryption.

\end{table}

In the past 15 years,  many cryptanalytic methods have been used to evaluate the security of MISTY1. For the low order degree of the S-boxes used in MISTY1,  Babbage\cite{BF} gave the first higher order differential cryptanalysis of 5-round MISTY1 without FL layers. Tsunoo et al.\cite{TSSK} introduced the higher order differential cryptanalysis of 7-round MISTY1 with FL layers, which is a chosen plaintext attack. K\"{u}hn \cite{K1} gave the first 6-round impossible differential cryptanalysis, which was improved by Lu et al.\cite{LKKD} with lower data complexity and time complexity. K\"{u}hn \cite{K2} introduced a slicing attack on 4 rounds MISTY1. Later, combining  the slicing attack and the generic impossible differential attack against 5-round Feistel constructions, Dunkelman et al.\cite{DK} gave a 6-round cryptanalytic result for MISTY1 with FL layers and a 7-round cryptanalytic result without FL layers.
Recently, taking advantage of some observations on FL functions and the early-abort technique, Jia et al.\cite{JL} improved a previous impossible differential attack on 6-round MISTY1 with 4 FL layers, 7-round MISTY1 with 3 FL layers and 7-round MISTY1 without FL layers. For the results in respect to the methods such as integral attacks, collision search attacks and  attacks under certain weak key assumptions, the related-key differential and amplified boomerang cryptanalysis of MISTY1, see \cite{SL},\cite{K1},\cite{DC},\cite{LKHLSHL},\cite{LYW}.

In this paper, we  apply the recent zero-correlation linear attacks to the block cipher MISTY1.
Zero-correlation linear cryptanalysis, proposed by Bogdanov and Rijmen\cite{BR}, is a novel promising attack technique for block ciphers.
It uses zero-correlation linear approximations generally existing in block ciphers to distinguish the differences between a
random permutation and a block cipher, which is different from the traditional linear cryptanalysis where high-correlation linear characteristics
are used. The initial distinguishers \cite{BR} have some limitations in terms of data complexity, which needs at least half of the codebook. Bogdanov and Wang \cite{BW} proposed a more data-efficient distinguisher by making use of  multiple zero-correlation linear approximations. The date complexity is reduced, however, the distinguishers rely on the assumption that all zero-correlation linear approximations are independent. To remove the unnecessary independency assumptions on the distinguishing side,  multidimensional distinguishers \cite{BLNW} were constructed for the zero-correlation property at AsiaCrypt 2012.  Recently, the multidimensional zero-correlation linear cryptanalysis were used in the analysis of the block cipher CAST-256\cite{BLNW}, CLEFIA\cite{BGWWC}, HIGHT\cite{WWBC}, E2\cite{WWB} and KASUMI\cite{YC}.

In this paper, we evaluate the security of MISTY1 with respect to the multidimensional zero-correlation linear cryptanalysis. Our contributions are summarized as follows:

1. We investigate the propagation characteristics of the linear masks on FL function and then propose four types of 5-round subkey-dependent zero-correlation linear approximation of MISTY1. Each type contains $2^{32}$ linear approximations, and the input masks and output masks of which are subkey-dependent. However, if  all $2^{32}$ linear approximations are taken in consideration in the key recovery process of MISTY1, there will be too many subkey bits involved that the time complexity will be greater than exhaustive search.
Fortunately, we find some observations on FO, FI and FL functions that when some special output masks are selected, only a part of keys influence the correlations of the linear approximations. For instance, if we set the right 9-bit output masks of FI function to zero, the absolute value of the correlations of the linear approximations will be independent with the involved subkeys. Those observations help us to select suitable linear approximations, which can reduce the number of the guessed subkey bits.

2. Based on the selected subkey-dependent zero-correlation linear approximations, we propose the multidimensional zero-correlation linear attack on 7-round MISTY1 with 4 FL layers using the partial-sum technique and the relations among subkeys obtained by the key schedule. Besides, MISTY1 without FL layers adopts strict balance Feistel structure, which has $2^{32}$ nature zero-correlation linear approximations\cite{BR}. We select $2^7$ ones based on the observations on FO, FI and FL functions, and propose the multidimensional zero-correlation linear attack on 7-round MISTY1 without FL layers

The paper is organized as follows. We list some notations,  briefly describe the block cipher MISTY1, outline the ideas of multidimensional zero-correlation linear cryptanalysis and explain why the subkey-dependent zero-correlation linear approximations are available in Section 2. Four types of 5-round subkey-dependent zero-correlation linear approximations of MISTY1 are shown in Section 3, as well as some observations on FL, FO and FI functions. Section 4 and Section 5 illustrate our attacks on 7-round  MISTY1 with 4 FL layers and without FL layers. We conclude this paper in Section 6.

\section{\large \bf  Preliminaries}
\subsection{\bf Notations}

%Throughout this paper, we will use some symbols, which are listed as follows:

\quad\ $FL_i$ \quad \quad :  the $i$-th FL function of MSITY1 with subkey $KL_i$.

$FO_i$ \quad \quad  :  the $i$-th FO function of MSITY1 with subkey $(KO_i, KI_i)$.

$FI_{ij}$ \quad \quad : the $j$-th FI function of $FO_i$ with subkey $KI_{i j}$.

$\wedge$ \quad \quad\quad\,: bitwise AND.

$\vee$ \quad\quad \quad\,: bitwise OR.

$\oplus$\quad\quad \quad\ : bitwise XOR.

$\neg$ \quad\quad \quad\,: bitwise NOT.

$a\cdot b$ \quad\quad \,: the scalar product of binary vectors by $a\cdot b = \oplus_{i=0}^{n-1}a_i b_i$.

%$a\diamond b$ \, \quad \, : the bitwise point multiplication of binary vectors by $a\diamond b =$
%
%
%\quad \,\quad \quad \,\quad \,$(a_0 b_0, a_1 b_1,..., a_{n-1} b_{n-1})$.

$X\|Y$ \ \,\quad: the concatenation of $X$ and $Y$.

$z[i]$ \quad \ \quad: the $i$-th bit of $z$, and $'0'$ is the most significant bit.

$z[i_1-i_2]$ : the $(i_2 - i_1 + 1)$ bits from the $i_1$-th bit to $i_2$-th bit of $z$.

$f\circ g$ \,\quad \ : the composite function of $f$ and $g$.

$f^{-1}$ \, \quad \ : the inverse function of $f$.

\subsection{\bf Description of KASUMI}
The MISTY1 algorithms \cite{M1} is a symmetric block cipher with a block size of 64 bits and a key size of 128 bits.
It adopts a 8-round Feistel structure with an FL layer every 2 rounds, see Fig. 1 (a) for an illustration. The FL layer consists of two FL functions. The FL function is a simple key-dependent boolean function, depicted in
Fig. 1 (d). Let the inputs of the FL function of the $i$-th round be $XL_i =
XL_{i,l}\|XL_{i,r}, KL_i =(KL_{i,1}, KL_{i,2})$, the output be $YL_i = YL_{i,l}\|YL_{i,r}$, where
$XL_{i,l}$, $XL_{i,r}$, $YL_{i,l}$ and $YL_{i,r}$ are 16-bit integers. We define the FL function as follows:
\begin{displaymath}
 \begin{array}{l}
YL_{i,r} =(XL_{i,l} \wedge KL_{i,1}) \oplus XL_{i,r};\\
YL_{i,l} =(YL_{i,r} \vee KL_{i,2})\oplus XL_{i,l}.
\end{array}
\end{displaymath}

The round function, that is the FO function, depicted in Fig. 1 (b), is another 3-round Feistel structure consisting of 3 FI functions and the key mixing stages. Let $XO_i = XO_{i,l}\|XO_{i,r}$, $KO_i =
(KO_{i,1},KO_{i,2},KO_{i,3}, KO_{i,4})$, $KI_i =(KI_{i,1}, KI_{i,2}, KI_{i,3})$ be the inputs of the $FO$
function of $i$-th round, and $YO_i = YO_{i,l}\|YO_{i,r}$ be the corresponding output,
where $XO_{i,l}$, $XO_{i,r}$, $YO_{i,l}$, $YO_{i,r}$ and $\overline{XI_{i,3}}$ are 16-bit integers. Then the FO function has the form
\begin{displaymath}
 \begin{array}{l}
\overline{XI_{i,3}} = FI((XO_{i,l}\oplus KO_{i,1}),KI_{i,1})\oplus XO_{i,r};\\
YO_{i,l} = FI((XO_{i,r} \oplus KO_{i,2}),KI_{i,2}) \oplus \overline{XI_{i,3}}\oplus KO_{i,4};\\
YO_{i,r} = FI((\overline{XI_{i,3}} \oplus KO_{i,3}),KI_{i,3})\oplus YO_{i,l}\oplus KO_{i,4}.
\end{array}
\end{displaymath}

\begin{figure}
  \centering
  \includegraphics[width=11cm]{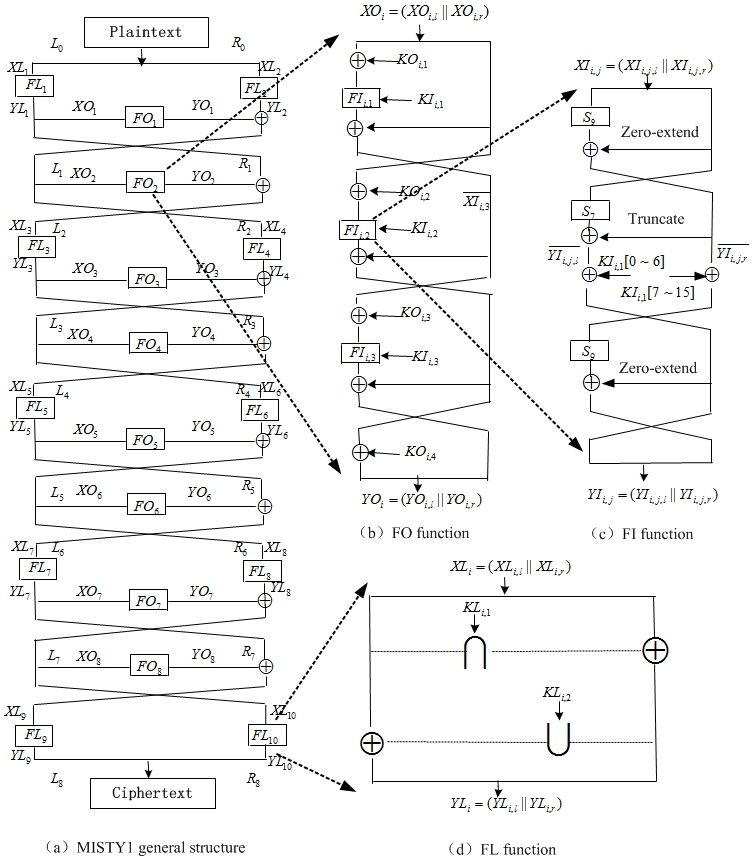}
  \caption{The structure and building blocks of MISTY1}
\end{figure}

 The structure of the FI function  is depicted in Fig. 1 (c). It uses two S-boxes $S_7$ and $S_9$ which are permutations of 7-bit to
7-bit and 9-bit to 9-bit respectively. Let $XI_{i,j}$, $YI_{i,j}$ be the inputs and the outputs of the $j$-th FI function
of the $i$-th round, where $XI_{i,j}$ and $YI_{i,j}$ are 16-bit integers. Denote that $KI_{i,j,l}=KI_{i,j}[0-8]$, $KI_{i,j,r}=KI_{i,j}[9-15]$, and $\overline{YI_{i,j,l}}$, $YI_{i,j,l}$ are two 7-bit variables, $\overline{YI_{i,j,r}}$, $YI_{i,j,r}$ are two 9-bit variables.
 The FI function can be described as follows:
\begin{displaymath}
 \begin{array}{l}
\overline{YI_{i,j,r}} = S_9(XI_{i,j}[0 - 8]) \oplus XI_{i,j} [9 -15];\\
\overline{YI_{i,j,l} } = S_7(XI_{i,j}[9-15]) \oplus \overline{YI_{i,j,r} };\\
YI_{i,j,l}  = \overline{YI_{i,j,l}}\oplus KI_{i,j,l};\\
YI_{i,j,r}  = S_9(\overline{YI_{i,j,r}}\oplus KI_{i,j,r})\oplus YI_{i,j,l};\\
YI_{i,j}  = YI_{i,j,l}\|YI_{i,j,r},
\end{array}
\end{displaymath}
where a 9-bit variable $\alpha$ bitwise XOR a 7-bit variable $\beta$, the 7-bit variable $\beta$ is first extend to 9-bit with two `0`bits in the left, while a 7-bit variable $\beta$ bitwise XOR a 9-bit variable $\alpha$, the 9-bit variable $\alpha$ is truncated to 7-bit ignoring the left two bits.

Let $(L_0,R_0)$, $(L_8,R_8)$ are the plaintext and ciphertext, and $L_i||R_i=\big((L_{i,l}\|L_{{i},r})\|(R_{i,l}\|R_{{i},r})\big)$ be the output of the $(i-1)$-th round, $i=1,2...,7$ and then the round function is defined as:

\begin{displaymath}
(L_i, R_i) = \left\{ \begin{array}{ll}
\Big(FO\big(FL(L_{i-1})\big)\oplus FL(R_{i-1}),FL(L_{i-1})\Big),\quad & \textrm{when }i =1, 3, 5, 7;\\
\Big(FO(L_{i-1})\oplus R_{i-1}, \,  L_{i-1}\Big), \quad & \textrm{when }i =2, 4, 6;\\
\Big(FL(FO(L_{i-1})\oplus R_{i-1}),\,FL(L_{i-1})\Big),\quad & \textrm{when }i =8.
\end{array} \right.
\end{displaymath}

The key schedule of MISTY1 takes the 128-bit key, which is divided into eight 16-bit words:
$(K_1,K_2, ..., K_8)$, that is $K =(K_1,K_2,...,K_8)$.  From this set of subkeys, another eight 16-bit words are generated according to $K'_i=FI_{K_{i+1}}(K_i)$. $KL_i = (KL_{i,1}, KL_{i,2})$, $KO_i = (KO_{i,1}, KO_{i,2}, KO_{i,3}, KO_{i,4})$,
and $KI_i = (KI_{i,1}, KI_{i,2}, KI_{i,3})$ are listed in Tab.2.
\begin{table}
\caption{The key schedule of MISTY1}
\centering
\scriptsize
\begin{tabular}{llllllllll}
\hline\noalign{\smallskip}
$KO_{i,1}$&$KO_{i,2}$ & $KO_{i,3}$ & $KO_{i,4}$ & $KI_{i,1}$ & $KI_{i,2}$ & $KI_{i,3}$ & $KL_{i,1}$ & $KL_{i,2}$ \\
\noalign{\smallskip}\hline\noalign{\smallskip}
$K_i$ &$K_{i+2}$ &$K_{i+7}$ &$K_{i+4}$&$K'_{i+5}$ &$K'_{i+1}$&$K'_{i+3}$& $K_{\frac{i+1}{2}}(\text{odd} \ i)$& $K'_{\frac{i+1}{2}+6}(\text{odd}\ i)$\\
      &       &           &           &            &      &      & $K'_{\frac{i}{2}+2}(\text{even}\ i)$& $K_{\frac{i}{2}+4}(\text{even}\ i)$\\
\noalign{\smallskip}\hline
\end{tabular}
\end{table}
\subsection{\bf Zero-correlation Linear cryptanalysis}

Consider a function $f : F^n_2 \mapsto F^m_2$ and let the input of the function be $x\in F_2^n$. A linear approximation
with an input mask $\alpha$ and an output mask $\beta$ is the following function:
$$x \mapsto \beta \cdot f(x)\oplus a\cdot x,$$
and its correlation  is defined as follows
$$C(\beta \cdot f(x), a\cdot x)=2Pr_{x}(\beta \cdot f(x)\oplus a\cdot x=0)-1.$$

In zero-correlation linear cryptanalysis, the distinguishers use linear approximations with zero correlation for
all keys while the classical linear cryptanalysis utilizes linear approximations with correlation far from zero. The basic idea of zero-correlation linear attack can be seen as the projection of impossible differential cryptanalysis to linear cryptanalysis. To construct the zero-correlation linear approximations, one adopts the miss-in-the-middle techniques just like to find impossible differential. Any linear approximations with nonzero bias is concatenated to any linear approximations with nonzero bias in the inverse direction, where the intermediate masks states contradict with each other.
%To reduce the data complexity and get rid of the dependent of the linear distinguishers,

Bogdanov et al. \cite{BW} proposed a multidimensional zero-correlation linear distinguisher using $\ell$ zero-correlation linear approximations and requiring $\mathcal{O}(2^{n+2}/\sqrt{\ell})$ distinct known plaintexts, where $n$ is the block size of a cipher. Let the zero-correlation linear distinguishers be the first $r-1$ rounds of a $r$-round block cipher. The zero-correlation linear approximations are available as a linear
space spanned by $m$ base zero-correlation linear approximations such that all
$\ell=2^m $ non-zero linear combinations of them have zero correlation. For
each of the $2^m$ data values $z \in F_2^m $, the attacker initializes a counter $V[z]$, $z=0, 1,...,2^m-1$ with all elements being zero.
Then, the attacker computes the data value $z$ in $F^m_2$ by evaluating the $m$ basis linear
approximations, that is, $z[i]=\alpha_i\cdot p \oplus \beta_i \cdot c'$ , $i=0,...,m-1$,
where  $(p, c')$ is obtained from
any distinct plaintext-ciphertext pair $(p, c)$ after partial decryption of the last round and $(\alpha_i,\beta_i)$ denotes the $i$-th basis linear approximation. Thus, the value of $z$ can be gotten, then the attacker increases the corresponding data value of the counter $V[z]$ by one. After all needed distinct plaintext-ciphertext pairs are computed, the attacker computes the statistic $T$:
$$T=\sum_{z=0}^{2^m-1}\frac{(V[z]-N2^{-m})^2}{N2^{-m}(1-2^{-m})}.\eqno(2.1)$$
The statistic $T$  follows a $\mathscr{X}^2$ -distribution with mean $\mu_0=(\ell-1)\frac{2^n-N}{2^n-1}$ and variance
$\sigma^2_0=2(\ell-1)\big(\frac{2^n-N}{2^n-1}\big)^2$ for the right key guess, while for the wrong key guess,
it follows a $\mathscr{X}^2$-distribution with mean $\mu_1=\ell-1$ and variance $\sigma_1^2=2(\ell-1)$.

If  the probability of the type-I error and the type-II error to distinguish between a wrong key and a right key are denoted as $\beta_0$ and $\beta_1$, respectively, considering the decision threshold $\tau =\mu_0+\sigma_0z_{1-\beta_0}=\mu_1-\sigma_{0}{z_{1-\beta_1}}$, the number of known plaintexts $N$ should be about

$$N=\frac{(2^n-1)(z_{1-\beta_0}+z_{1-\beta_1})}{\sqrt{(\ell-1)/2}+z_{1-\beta_0}}+1,\eqno(2.2)$$
where $z_{1-\beta_0}$ and $z_{1-\beta_1}$are the respective quantiles of the standard normal distribution. More
details are described in \cite{BLNW}.

However, we employ $\ell=2^m $ non-zero linear combinations of the $m$ base subkey-dependent zero-correlation linear approximations, that is, the output masks of the linear approximations can be deduced by the input masks and the subkeys. When the guessed subkeys are right, $T$ follows a $\mathscr{X}^2$ -distribution with mean $\mu_0=(\ell-1)\frac{2^n-N}{2^n-1}$ and variance $\sigma^2_0=2(\ell-1)\big(\frac{2^n-N}{2^n-1}\big)^2$, while the guessed subkeys are wrong, $V[z]$ also follows the multinomial distribution,  although the linear approximations used can not be guaranteed to be zero-correlation linear approximations, that is, for the wrong key guess, $T$ also follows a $\mathscr{X}^2$-distribution with mean $\mu_1=\ell-1$ and variance $\sigma_1^2=2(\ell-1)$.

\begin{figure}
\centering
\includegraphics[width=10cm]{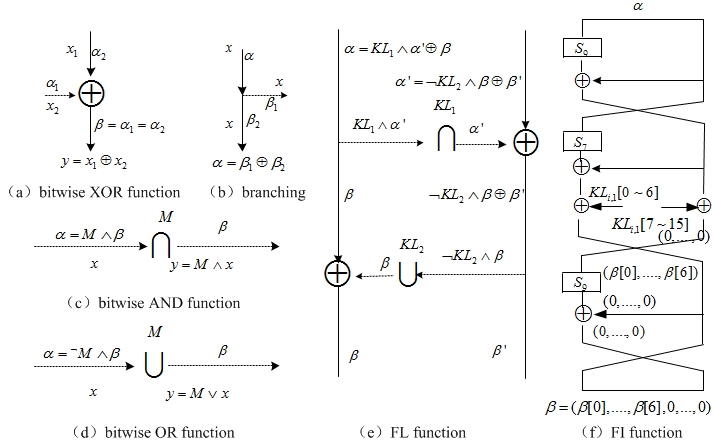}
\caption{Property of  XOR, Branching, AND, OR, FL and FI functions}
\label{fig:2}
\end{figure}

Thus, the multidimensional zero-correlation linear cryptanalysis using the subkey-dependent linear approximations can be processed in the following steps.
\begin{itemize}
\vspace{-0.1in}\item[\rm (1)]  Find some longest subkeys-dependent zero-correlation linear approximations by using  properties of encryption algorithm. Assume that the dimensional number of the distinguishers is $\ell$.
\vspace{-0.1in}\item[\rm (2)]  Allocate a subkey-dependent counter $V_k[z]$ for $\ell$-bit $z$. The vector $z$ is the concatenation of evaluations of $\ell$ subkey-dependent zero-correlation linear approximations.
\vspace{-0.1in}\item[\rm (3)]  Update the counter $V_k[z]$ by guessing the subkeys in the process of partial-encryption and partial-decryption one after another by using the partial-sum technique.
\vspace{-0.1in}\item[\rm (4)] Guess the linear approximations involved subkeys $k$ , compute statistic $T$ according Equation (2.1). If $T\leq \tau$, then the guessed subkey values are possible right subkey candidates, where $\tau$ is computed in traditional method.
\vspace{-0.1in} \item[\rm (5)] Do exhaustive search for all right candidates.
\end{itemize}

Furthermore, the relations among subkeys can be obtained by
carefully analyzing the key schedule algorithm, which can be used to reduce the subkey bits guessed.

\section{\large\bf Some Observations in MSITY1 }
In this section, we propose some zero-correlation linear approximations of MISTY1. Moreover, some properties of FL, FI and FO functions are discovered.
\begin{lem}$^\text{\cite{BR}, \cite{YC}}$
Let $M$ be a $\ell$-bit value and define the XOR, Branching, OR, AND functions $h_1$, $h_2$, $h_3$ and $h_4$ as $h_1(x_1,x_2)=x_1 \oplus x_2$, $h_2(x)=(x,x)$, $h_3(x)=M \vee x$, $h_4(x)=M \wedge x$, see Fig.2 (a,b,c,d). Then we have

\begin{itemize}
\item[\rm (1)]  For any masks $\alpha_1$, $\alpha_2$ and $\beta$, $C\big(\beta \cdot h_1(x_1,x_2),(\alpha_1,\alpha_2)\cdot (x_1,x_2)\big)\neq 0$ if and only if $ \beta= \alpha_1=\alpha_2$;
\item[\rm (2)]  For any  masks $\alpha$, $\beta_1$ and $\beta_2$, $C\big((\beta_1, \beta_2) \cdot h_2(x),\alpha\cdot x\big)\neq 0$ if and only if $ \alpha= \beta_1 \oplus \beta_2$;
\item[\rm (3)]  For any $\ell$-bit masks $\alpha$ and $\beta$, $C(\beta \cdot h_3(x),\alpha\cdot x)\neq 0$ if and only if $ \alpha= \urcorner M\wedge\beta$;
\item[\rm (4)] For any $\ell$-bit masks $\alpha$ and $\beta$, $C(\beta\cdot h_4(x),\alpha\cdot x)\neq 0$ if and only if $ \alpha= M\wedge \beta$.
\end{itemize}
\end{lem}
%Base on Observation 1, we have the following  result about the $FI$ function.

\begin{lem}
Let  $(\alpha, \alpha' )$, $(\beta, \beta' )$ be the input and the corresponding output masks of the linear function $FL_i$ with nonzero-correlation, then for any $0\leq j\leq 15$, we have
$$\alpha'[j]=\neg KL_{i,2}[j]\beta[j]\oplus \beta'[j],\ \text{and}\  \alpha[j]= KL_{i,1}[j]\alpha'[j]\oplus \beta[j],$$
 which can be denote by $(\alpha[j], \alpha'[j])=\overline{FL_i}^j(\beta[j], \beta'[j]; KL_{i,1}[j], KL_{i,2}[j])$, see Fig.2(e) for detail.
 Then the function $\overline{FL_i}=(\overline{FL_{i}}^0,...,\overline{FL_{i}}^{15})$ has the following two properties:
\begin{itemize}
\item[\rm (1)]
The function $\overline{FL_i}$ can be denoted as
$$(\alpha, \alpha' )=\overline{FL_i}(\beta, \beta'; KL_{i,1}, KL_{i,2} ),$$  and $(\alpha'[j],\alpha[j])$ are not influenced by $(\beta'[k], \beta[k])$ and $(KL_{i,1}[k], KL_{i,2}[k])$, when $j\neq k$.
\item[\rm (2)]The function $\overline{FL_i}$ is linear, that is $$\overline{FL_i}(\beta_1\oplus \beta_2)=\overline{FL_i}(\beta_1)\oplus \overline{FL_i}(\beta_2),$$
     which means that the $\overline{FL_i}$ function is  a linear components for the masks.
\end{itemize}

\end{lem}

\begin{lem}
Let  $\alpha$, $\beta$ be the 16-bit input and the corresponding output masks of the  function $FI_{i,j}$, where $0\leq i\leq 7$ and $0\leq j \leq 2$, then $\beta=0$ leads to $\alpha=0$, if the correlation of the linear approximation is nonzero.
\end{lem}

\proof   Form Lemma 3 in \cite{BR}, we only need to know the function FI is invertible, which can be proved easily and we omit it for simplicity.

With those three lemmas, we have some  zero-correlation linear approximations on 5-round MISTY1. We give the descriptions in the following theorem.

\begin{thm}
Let $(0, \beta)$ be the output masks of round 7 and $(\alpha, 0)$ be the input masks of round 3, then
$(\alpha, 0)\overset{\text{5-Round}}{\longrightarrow}(0,\beta)$
is a zero-correlation linear approximations form round 3 to 7, when $\alpha=\overline{FL_3}\circ\overline{FL_5}\circ\overline{FL_7}(\beta)$ and $\beta$ is a 32-bit nonzero value, see Fig. 3(a). For the MISTY1 without FL functions, $(\beta, 0)\overset{\text{5-Round}}{\longrightarrow}(0,\beta)$ is a zero-correlation linear approximations form round $r$ to $r+4$, where $\beta$ is a 32-bit nonzero value and $r=1,2,3$, see Fig. 3(b).
\end{thm}

\begin{figure}
\centering
\includegraphics[width=8cm]{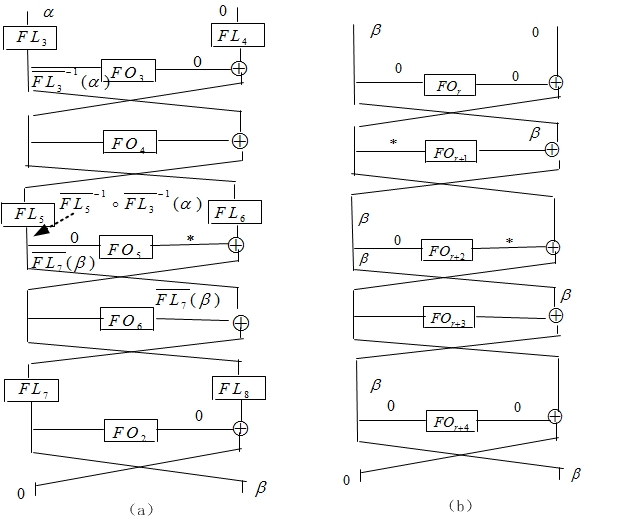}
\caption{Zero-correlation linear approximations on 5-round MISTY1}
\label{fig:3}
\end{figure}

%Let the input masks of round 2 be  $(\alpha, 0)$, and the output masks of round 6 be
%$(0, \beta)$,  we deduce that the output masks of $FO_4$ is nonzero, and the input masks of $FO_4$ is $\overline{FL_8}\circ\overline{FL_6}(\beta)\oplus\overline{FL_4}(\alpha)=0$. Hence $(\alpha, 0)\overset{\text{5-Round}}{\longrightarrow}(0,\beta)$
%is an linear approximations with correlation zero for rounds 2 to 6, see Fig. 3(a). We also have the other three types of   5-round zero-correlation subkeys-dependent linear approximations, see Fig. 3(b,c,d).

Some properties of FI and FO functions can be described as the following two observations.

\begin{obs}
Let $\beta$ be a $16$-bit value with $\beta[7-15]=0$ and $YFI_{i,j}$ the 16-bit output of the $FI_{i,j}$ function. Then, the value of $\beta \cdot YFI_{i,j}$ can be written as $X \oplus (\beta[0-6] \cdot KI_{i,j}[0-6])$, where $X$ is an value independent with $KI_{i,j}$.
\end{obs}

\begin{obs}
 Let $ (\beta, \beta')$ be a 32-bit value, such that $\beta[7-15]=\beta'[7-15]=0$, and $YFO_{i}$ the 32-bit output of the $FO_{i}$ function.
 Then the value of $(\beta,\beta') \cdot YFO_{i}$ can be written as $Y \oplus\big (\beta[0-6] \cdot (KI_{i,3}\oplus KI_{i,2}\oplus KO_{i,4})[0-6]\big) \oplus (\beta[0-6] \cdot KI_{i,2}[0-6])$, where $Y$ is an value dependent with $KO_{i,1}$, $KO_{i,2}$, $KO_{i,3}$ and $KI_{i,1}$
\end{obs}

\section{\large\bf Key-Recovery Attack on 7-Round MISTY1 with 4 FL Layers }

In this section, we extend our attacks to 7-round(2-8) MISTY1 with 4 FL Layers. We mount the 5-round zero-correlation linear approximations from round 3 to 7, and extend one round forward and one round as well as one FL layer backward, respectively, see Fig.4(c).
We select the 5-round zero-correlation linear approximations as $(\alpha_1\|\alpha_2, 0)\overset{\text{2 to 6 round}}{\longrightarrow}(0,\beta\|0),$ that is
$$(\alpha_1,\alpha_2)\cdot (L_{1,l},L_{1,r}) \oplus (\beta, 0) \cdot (R_{8,l},R_{8,r})=0,$$
where $\beta$ is 16-bit non-zero value with  $\beta[7-15]=0$, and $\alpha_1$, $\alpha_2$ are two 16-bit non-zero values with $\alpha_1\|\alpha_2=\overline{FL_3}\circ\overline{FL_5}\circ\overline{FL_7}(\beta\|0)$. Then, by Lemma 2, we can know that $\alpha_1[7-15]=\alpha_2[7-15]=0$.

\textbf{Basis Subkey-dependent Zero-correlation Masks.}  From the used subkey-dependent zero-correlation linear approximates, we know that $\alpha_1$, $\alpha_2$ are two 16-bit non-zero values with $\alpha_1\|\alpha_2=\overline{FL_3}\circ\overline{FL_5}\circ\overline{FL_7}(\beta\|0)$, then 42-bit keys $K_2[0-6], K_3[0-6], K_4[0-6], K'_8[0-6], K'_1[0-6], K'_2[0-6]$ are involved. Assume that $m_0$,...,$m_6$ are seven 32-bit values, and $m_0=(1,0,..,0)$, $m_1=(0,1,0,...,0)$,...,$m_6=(0,0,0,0,0,0,1,0,...,0)$.
For $2^{42}$ possible involved subkeys, compute
 $$(\alpha_1)_i\|(\alpha_2)_i=\overline{FL_3}\circ\overline{FL_5}\circ\overline{FL_7}(m_i),$$
Store the value $z_i=(m_i[0-6]\| (\alpha_1)_i[0-6]\|(\alpha_2)_i[0-6])$, $i=0,1,...,6$ in a hash table $T_1$ indexed by the 42-bit involved subkeys.

\textbf{Attack Process.} The key-recovery attacks on  7-round MISTY1 with 4 FL layers are proceeded with the partial-sum technique as follows.
\begin{figure}
\centering
  \includegraphics[width=10cm]{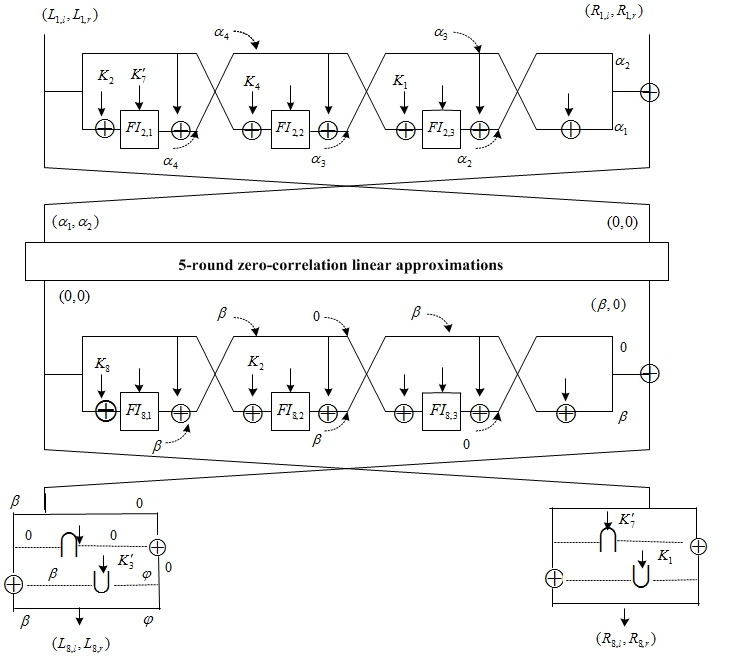}
  \caption{ Attacks on 7-round MISTY1 with 4 FL layers }
  \label{fig:5}
\end{figure}

1. Collect $N$ plaintexts with corresponding ciphertexs.  Allocate a 8-bit counter $V_0[y_0]$ for each of $2^{92}$ possible values of $$y_0=y_0^1\|y_0^2\|y_0^3\|y_0^4\|y_0^5\|y_0^6\|y_0^7\|y_0^8,$$
where $y_0^1 =L_{1,l}$, $y_0^2=L_{1,r}$,
$y_0^3 =R_{1,l}[0-6]$, $y_0^4=R_{1,r}[0-6]$,
$y_0^5 =R_{8,l}$, $y_0^6=R_{8,r}$, $y_0^7 =L_{8,l}[0-6]$, $y_0^8=L_{8,r}[0-6]$, and set them zero. Calculate the
number of pairs of plaintext-ciphertext with given values $y_0$ and save it in $V_0[y_0]$. In
this step, around $2^{64}$ plaintext-ciphertext pairs are divided into $2^{92}$ different states.  So the assumption $V_0$ as a 8-bit counter is sufficient.

2. Allocate a counter $V_1[y_1]$ for each of $2^{60}$ possible values of $$ y_1= y_1^1\|y_1^2\|y_1^3\|y_1^4\|y_1^5\|y_1^6,$$ where $y_1^1=y_0^2$, $y_1^2=y_0^8$,
 and set them zero. For all $2^{69} $ possible values of $y_0^1$, $y_0^3$, $y_0^4$, $y_0^5$, $y_0^6$ and $y_0^8$, guess the 48-bit $KO_{2,1}$, $KL_{10,1}$, $KL_{10,2}$, that is $K_2$, $K'_7$ and $K_1$, compute
 $$y_1^3= y_0^3\oplus \big(FI_{2,1}(y_0^2\oplus K_2, K'_7) \oplus y_1^1\big)[0-6];$$
 \begin{displaymath}
\begin{array}{l}
y_1^4= y_0^4\oplus S_9\big(\big(FI_{2,1}(y_0^2\oplus K_2, K'_7) \oplus y_1^1\oplus K_1\big)[0-8]\big)\\
\qquad\quad\ \, \oplus\big(FI_{2,1}(y_0^2\oplus K_2, K'_7) \oplus y_1^1\oplus K_1\big)[9-15]\\
\qquad\quad\ \, \oplus S_7\big(\big(FI_{2,1}(y_0^2\oplus K_2, K'_7) \oplus y_1^1\oplus K_1\big)[9-15]\big);
\end{array}
\end{displaymath}
 $$y_1^5= (y_0^6\vee K_1) \oplus y_0^5;$$
 \begin{displaymath}
\begin{array}{l}
y_1^6= y_0^7\oplus (y_1^5\wedge K'_7)\oplus y_0^6\oplus
 S_9\big(\big((y_1^5\wedge K'_7)\oplus y_0^6\oplus K_2\big)[0-8]\big)\\
 \qquad\quad\ \,\oplus\big((y_1^5\wedge K'_7)\oplus y_0^6\oplus K_2\big)[9-15]\\
\qquad\quad\ \, \oplus S_7\big(\big((y_1^5\wedge K'_7)\oplus y_0^6\oplus K_2\big)[9-15]\big),
\end{array}
\end{displaymath}
and update the value $V_1[y_1] = V_1[y_1] + V_0[y_0]$.

3.  Allocate a counter $V_2[y_2]$ for each of $2^{53}$ possible values of
$$y_2=y_2^1\|y_2^2\|y_2^3\|y_2^4\|y_2^5,$$
 where $y_2^1 = y_1^1$, $y_2^2= y_1^3$, $y_2^3 = y_1^4$, $y_2^4 = y_1^5$, and set them zero.
For all $2^{7} $ possible values of $y_1^2$, guess the 7-bit $KL_{9,2}[0-6]$, that $K'_3[0-6]$, and compute
$$y_2^5= y_1^6 \oplus( K'_3[0-6]\wedge y_1^2[0-6]),$$
and update the value $V_2[y_2] = V_2[y_2] + V_1[y_1]$.

4. Allocate a counter $V_3[y_3]$ for each of $2^{44}$ possible values of
$$y_3=y_3^1\|y_3^2\|y_3^3\|y_3^4\|y_3^5,$$ where
 $y_3^1= y_2^1$, $y_3^2 = y_2^2$, $y_3^3 = y_2^3$, $y_3^4 = y_2^4[9-15]$ and set them zero. For all $2^{9} $ possible values of $y_2^4[0-8]$, guess the 9-bit $KO_{8,1}[0-8]$, that is $K_8[0-8]$, compute
 $$y_3^5= y_2^5 \oplus S_9(y_2^4[0-8]\oplus K_8[0-8])[2-8],$$
  and update the value $V_3[y_3] = V_3[y_3] + V_2[y_2]$.

5. Allocate a counter $V_4[y_4]$ for each of $2^{37}$ possible values of
 $$y_4=y_4^1\|y_4^2\|y_4^3\|y_4^4,$$
 where $y_4^1 = y_3^1$, $y_4^2 = y_3^2$, $y_4^3 = y_3^3$,
 and set them zero. For all $2^{7} $ possible values of $y_3^4$, guess the 7-bit $KO_{8,1}[9-15]$, that is $K_8[9-15]$, compute
 $$y_4^4= y_3^5 \oplus y_3^4[0-6]\oplus S_7(y_3^4[0-6]\oplus K_4[9-15]),$$ and update the value $V_4[y_4] = V_4[y_4] + V_3[y_3]$.

6. Allocate a counter $V_5[y_5]$ for each of $2^{28}$ possible values of
$$y_3=y_5^1\|y_5^2\|y_5^3\|y_5^4,$$
 where $y_5^1=y_4^1[9-15]$, $y_5^4=y_4^4$  and set them zero.
 For all $2^{9} $ possible values of $y_4^1[0-8]$, guess the 9-bit $KO_{2,2}[0-8]$, that is $K_4[0-8]$, compute
  $$y_5^2= y_4^2 \oplus S_9(y_4^1[0-8]\oplus K_8[0-8])[2-8];$$
  $$y_5^3= y_4^3 \oplus S_9(y_4^1[0-8]\oplus K_8[0-8])[2-8],$$
and update the value $V_5[y_5] = V_5[y_5] + V_4[y_4]$.

7. Allocate a counter $V_6[y_6]$ for each of $2^{21}$ possible values of
 $$y_6=y_6^1\|y_6^2\|y_6^3,$$
 where $y_6^3 = y_5^4$, and set them zero. For all $2^{7} $ possible values of $y_5^1$, guess the 7-bit $KO_{2,2}[9-15]$, that is $K_4[9-15]$, compute $$y_6^1= y_5^2 \oplus y_5^1\oplus S_7(y_5^1\oplus K_4[9-15]);$$
 $$y_6^2= y_5^3 \oplus y_5^1\oplus S_7(y_5^1\oplus K_4[9-15]),$$
 and update the value $V_6[y_6] = V_6[y_6] + V_5[y_5]$.

8. Up to now, we have guessed 87-bit subkeys $K_1$, $K_2$, $K'_3[0-6]$, $K_4$, $K'_7$ and $K_8$ and we can deduce $K'_8$ from $K_8$ and $K_1$ by the key schedule. Guess the 9-bit $K'_3[7-15]$, and we can deduce $K_3$ from $K'_3$ and $K_4$, deduce $K'_2$ from $K_3$ and $K_2$. Form table $T_1$, we can get the 7 basic subkey-dependent zero-correlation masks $z_i$, $i=0,1,...,6$.

9. Allocate a 64-bit counter vector $V[z]$ for 7-bit $z$, where $z$ is the concatenation of evaluations of 7 basis subkey-dependent zero-correlation masks $z_i$, $i=0,1,...,6$. Compute $z$ from $y_6$ with 7 subkey-dependent basis zero-correlation masks $z_i$, $i=0,1,...,6$, save it in $V[z]$, that is $V [z]+ = V_6[y_6]$.

10. Compute the statistic $T$ according to Equation (2.1). If $T<\tau$ , the guessed key value is a right key candidate.
As there are 32 master key bits that we have not guessed, we do exhaustive search for all keys conforming to this possible key candidate.

\textbf{Complexity of the Attack.}
In this attack, we set the type-I error probability $\beta_0 =2^{-2.7}$ and the type-II error probability $\beta_1 =2^{-10}$. We have
$z_{1-\beta_0}\approx 1$, $z_{1-\beta_1}\approx 3.09$, $n = 64$, $ l=2^{7}$. The date complex $N$ is about $2^{62.9}$ and the decision threshold $\tau \approx 2^{6.23}$. The time complexity of steps 1-10 in the described attack is as follows:
\begin{itemize}
\item[\rm (1)] Step 1 requires $2^{62.9}$ memory accesses.
\item[\rm (2)] Step 2 requires $2^{62.9}\times 2^{48}$ memory accesses, because we should guess 48 bits $K_1$, $K_2$ and $K'_7$,
               and compute $y_1$ from $y_0$, and then  update $V_1$ for $2^{62.9}$ times.
\item[\rm (3)] Step 3 requires $2^{48}\times 2^{7}\times 2^{60}$ memory accesses, because for all of guessed 48 bits
               keys in previous steps we should guess 7 bits for $K'_3[0-6]$, and compute $y_2$ from $y_1$, and then update $V_2$  for $2^{60}$ times.
\item[\rm (4)] Step 4 requires $2^{55}\times 2^{9}\times 2^{53}$ memory accesses, because for all of guessed 55 bits
               keys in previous steps we should guess 9 bits for $K_8[0-8]$, and compute $y_3$ from $y_2$, and then update $V_3$  for $2^{53}$ times.
\item[\rm (5)] Step 5 requires $2^{64}\times 2^{7}\times 2^{44}$ memory accesses, because for all of guessed 64 bits
               keys in previous steps we should guess 7 bits for $K_8[9-15]$, and compute $y_4$ from $y_3$, and then update $V_4$  for $2^{44}$ times.
\item[\rm (6)] Step 6 requires $2^{71}\times 2^{9}\times 2^{37}$ memory accesses, because for all of guessed 71 bits
               keys in previous steps we should guess 9 bits for $K_4[0-8]$, and compute $y_5$ from $y_4$, and then update $V_5$  for $2^{37}$ times.
\item[\rm (7)] Step 7 requires $2^{80}\times 2^{7}\times 2^{28}$ memory accesses, because for all of guessed 80 bits
               keys in previous steps we should guess 7 bits for $K_4[9-15]$, and compute $y_6$ from $y_5$, and then update $V_6$  for $2^{28}$ times.
\item[\rm (8)] Step 8 requires $2^{42}\times 7$ Table $T_1$ accesses.
\item[\rm (9)] Step 9 requires $2^{96}\times 2^{21}$ memory accesses.
\item[\rm (10)] Step 10 requires $2^{86}\times 2^{32}$ 7-round MISTY1 encryption. Because $2^{96}\cdot 2^{-10} = 2^{86}$ key candidates can survive in the wrong key filtration.
\end{itemize}
 If we consider one memory accesses as a 7-round encryption, the total time complexity is about $2^{119.5}$ of 7-round MISTY1. The total compute complexity is about $2^{119.5}$ 7-round MISTY1 encryptions with $2^{62.9}$ known plaintexts and $2^{93}$ memory bytes for counters.

\section{\bf Key-Recovery Attack on 7-Round  MISTY1 without FL Layers}

In this section, the multidimensional zero-correlation linear attack on 7-round MISTY1 without FL layers is presented. The 5-round zero-correlation linear approximations start from 3 round and end at 7 round,
and extend one round forward and backward, respectively, see Fig. 5. We select the 5-round zero-correlation linear approximations as:
$$(\beta\|0, 0)\overset{\text{ 3 to 7 round }}{\longrightarrow}(0,\beta\|0),$$
where $\beta$ is 16-bit non-zero value with  $\beta[7-15]=0$. The choice is to minimize the
key words guessing during the attack on 7-round  MISTY1. Form Observations 3, we know that, $KI_{2,1}$, $KI_{2,2}$, $KI_{2,3}$, $KO_{2,3}$,$KO_{2,4}$, $KI_{8,1}$, $KI_{8,2}$, $KI_{8,3}$, $KO_{8,3}$ and $KO_{8,4}$ are not involved in the computation, which can help us to reduce the complexity of the attack.
\begin{figure}
\centering
  \includegraphics[width=10cm]{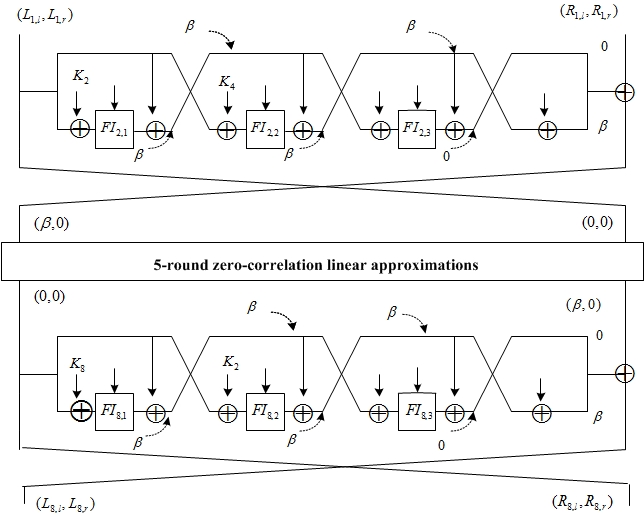}
  \caption{Attacks on 7-round MISTY1 without FL layers}
  \label{fig:4}
\end{figure}

\textbf{Basis Zero-correlation Masks.} Assume that $m_0$,...,$m_6$ are seven 16-bit values, and $m_0=(1,0,..,0)$, $m_1=(0,1,0,...,0)$,..., $m_6=(0,0,0,0,0,0,1,0,...,0)$. Let $z_i=(m_i[0-6]\|m_i[0-6])$, $i=0,1,...,6$, then we know $\{z_i\}_{i=0,1,...,6}$ are a group of basic zero-correlation masks.

\textbf{Attack Process.} The zero-correlation linear attack on  7-round MISTY1 without FL layers with the partial-sum technique is demonstrated as follows.

%In this section, we describe our attacks on 6 rounds of KASUMI. We use the FFT to reduce time complex in our attack. The number of guessed key bits is affected by several parameters including the zero-correlation linear property we choose (values of ¦Á and ¦Â), the position of the property(rounds spanned by zero-correlation approximations), and
%
%the number of rounds added before and after this property. To optimize the attack complexities, a proper choice
%of these parameters is needed. We have implemented the
%search for the best parameters in a computer program
%which counts the number of guessed key bits in the par-
%tial encryption/decryption phase for all possible combina-
%tions of the parameters. To reduce the time complexity, we
%choose parameters with the least number of guessed key
%bits.

%In our attack, we guess the subkeys and evaluate the linear approximation $(\beta,\beta')^{T}\cdot \big((L_{0,l}\oplus L_{6,l}), (L_{0,r} \oplus L_{6,r})\big)=0$, that is
%$$(\beta,\beta')\cdot(L_{0,l}\oplus L_{6,l}\oplus L_{5,r},L_{0,r}\oplus L_{6,r})\oplus \beta\cdot\big(FI(L_{5,l}\oplus( k_1\lll 5), k'_4) \oplus FI(L_{5,r}\oplus (k_5\lll8), k'_3)\big)=0,$$
%where  $\beta'=(\beta\ggg 1)\diamond \urcorner KL_{6,2}$ and  $\beta$ is any 16-bit non-zero value with $\beta[0-6]=\beta[7-13]$ and $\beta[14]=\beta[15]=0$.
% Then the key recovery attack on 6-round KASUMI is proceeded with Partial-sum technique as follows:

1. Collect $N$ plaintexts with corresponding ciphertexs.  Allocate a 8-bit counter $N_0[x_0]$ for each of $2^{71}$ possible values
$$x_0=x_0^1\|x_0^2\|x_0^3\|x_0^4\|x_0^5,$$
where $x_0^1 =L_{1,r}$, $x_0^2=L_{1,l}$, $x_0^3=R_{8,r}$, $x_0^4=R_{8,l}$, and $$x_0^5=R_{1,r}[0-6]\oplus R_{8,r}[0-6]\oplus L_{1,r}[0-6]\oplus R_{8,r}[0-6],$$ and set them zero. Calculate the
number of pairs of plaintext-ciphertext with given values $x_0$ and save it in $N_0[x_0]$. In
this step, around $2^{64}$ plaintext-ciphertext pairs are divided into $2^{71}$ different states. So the assumption $N_0$ as a 8-bit counter is sufficient.

2. Allocate a counter $N_1[x_1]$ for each of $2^{39}$ possible values
$$x_1=x_1^1\|x_1^2\|x_1^3,$$
where $x_1^1=x_0^1$, $x_1^2=x_0^3$, and set them zero.  For $2^{32} $ possible values of $x_{0}^2\|x_{0}^4$, guess the 16-bit $KO_{1,1}=KO_{8,2}$, that is $K_2$, and compute
\begin{displaymath}
\begin{array}{l}
x_1^3= x_0^5\oplus S_9\big((x_0^2\oplus K_2)[0-8]\big)\oplus(x_0^2\oplus K_2)[9-15]\oplus
S_7\big((x_0^2\oplus K_2)[9-15]\big)\\
\quad \quad \quad \ \ \oplus S_9\big((x_0^4\oplus K_2)[0-8]\big)\oplus(x_0^4\oplus K_2)[9-15]\oplus S_7\big((x_0^4\oplus K_2)[9-15]\big),
\end{array}
\end{displaymath}
and update the value $N_1[x_1] = N_1[x_1] + N_0[x_0]$.

3. Allocate a counter $N_2[x_2]$ for each of $2^{23}$ possible values of
$$x_2=x_2^1\|x_2^2,$$
 where $x_2^1 = x_1^2$ and set them zero. For all $2^{16} $ possible values of $x_1^1$, guess the 16-bit $KO_{1,2}$, that is $K_4$, and
compute $$x_2^2=x_1^3\oplus S_9\big((x_1^1\oplus K_4)[0-8]\big)\oplus(x_1^1\oplus K_4)[9-15]\oplus
S_7\big((x_1^1\oplus K_4)[9-15]\big),$$
and update the value $N_2[x_2] = N_2[x_2] + N_1[x_1]$.

4. Allocate a counter $N_3[x_3]$ for each of $2^{7}$ possible values of
$x_3$, and set them zero. For all $2^{16} $ possible values of $x_2^1$, guess the 16-bit $KO_{8,1}$, that is $K_8$, and compute
$$x_3=x_2^2\oplus S_9\big((x_2^1\oplus K_8)[0-8]\big)\oplus(x_2^1\oplus K_8)[9-15]\oplus
S_7\big((x_2^1\oplus K_8)[9-15]\big),$$
and then, update the value $N_3[x_3] = N_3[x_3] + N_2[x_2]$.

5.  Allocate a 64-bit counter vector $N[z]$ for 7-bit $z$, where $z$ is the concatenation of evaluations of 7 basis zero-correlation masks $z_i$, $i=0,1,...,6$. Compute $z$ from $x_3$ with 7 basis zero-correlation masks, save it in $N[z]$, that is $N [z]+ = N_3[x_3]$.

6. Compute the statistic $T$ according to Equation (2.1). If $T<\tau$ , the guessed key value is a right key candidate. As there are 80 master key bits that we have not guessed, we do exhaustive search for all keys conforming to this possible key candidate.

\textbf{Complexity of the Attack.} In this attack, we set the type-I error probability $\beta_0 =2^{-2.7}$ and the type-II error probability $\beta_1 =2^{-48}$. We have
$z_{1-\beta_0}\approx 1$, $z_{1-\beta_1}\approx 7.91$, $n = 64$, $ l=2^{7}$. The date complex $N$ is about $2^{63.9}$ and the decision threshold $\tau \approx 2^{6.97}$.

During the encryption and decryption phase,  the complexity of Step 1,2,3,4 is no more than $2^{63.9}$ memory access, $2^{63.9}\times2^{16}$ memory access, $2^{39}\times 2^{16} \times 2^{16}=2^{71}$ memory access and $2^{23}\times 2^{16} \times2^{16} \times2^{16} =2^{71}$ memory access, respectively.

Step 5 can be done independently. The complexity of Step 5 is no more than $7\times 2^{7}\times 2^{48}$ memory access.

Step 6 requires $2^{80}$ 7-round MISTY1 encryption, because for all of guessed $2^{48}$ keys in previous steps,
only the right key candidate can survive in the wrong key filtration, the complexity of Step 7 is about $2^{80}$ 7-round MISTY1 encryption.

If we consider one memory accesses as a 7-round encryption, the total time complexity is about $2^{118}$ of 6-round MISTY1.  In total, the data complexity is about $2^{63.9}$ known plaintexts, the time complexity is about $2^{81}$ of 7-round encryptions and the memory requirements are $2^{61}$ bytes for counters.

\section{\large\bf Conclusion }
In this paper, we have investigated the security of 7-round MISTY1 with 4 FL layers by means of multidimensional zero-correlation linear cryptanalysis. Firstly, some  properties of the FL function in MISTY1 have been proposed, following which we have constructed the
first known 5-round subkey-dependent zero-correlation linear distinguishers of MISTY1 with FL layers. This distinguishers cover the same number of rounds as the zero-correlation linear distinguishers for MISTY1 without FL layers and we explain that those subkey-dependent zero-correlation linear distinguishers can also be used in the multidimensional zero-correlation linear cryptanalysis. In order to reduce the number of guessed subkey bits, we find some observations on FO, FI and FL functions and select out a part of the subkey-dependent zero-correlation linear approximations and then, conduct the multidimensional zero-correlation attacks on the 7-round MISTY1 with 4 FL layers. The advantages of the attack over exhaustive search
are about $128-119.5=8.5$ bits. Besides, multidimensional zero-correlation linear cryptanalysis of 7-round MISTY1 without FL layers has been conducted. None of the two attacks directly threatens the security of MISTY1, but they reduce the security margin of the cipher and give us an example to evaluate the effects of the FL layers.

\end{document}